\documentclass[12pt]{iopart}

\usepackage{graphicx}
\begin{document}

\title[]{Cosmological Solutions in $2+1$ Dimensional New Massive Gravity in the presence of the Dirac field}

\author{Ganim Gecim and Yusuf Sucu}

\address{Department of Physics, Faculty of Science,
Akdeniz University, \\ 07058 Antalya, Turkey}
\ead{gecimganim@gmail.com and ysucu@akdeniz.edu.tr} \vspace{10pt}

\begin{abstract}
In this paper, we consider $2+1$ dimensional gravitational theory
including a Dirac field that is minimally coupled to New Massive Gravity. We investigate cosmological solutions of the field
equations by using the self-interaction potential form obtained by the existence of Noether
symmetry. In this context, we obtain cosmological solutions that corresponds to inflationary as well as
the oscillationary epochs of the universe. Moreover, we have seen that the Dirac field behaves like a dark energy in these epochs of the universe.
\end{abstract}


\maketitle

\section{Introduction}

\label{sec:intro}

New Massive Gravity (NMG)  proposed by Bergshoeff, Hohm and Townsend
by adding a specific quadratic curvature term to the
Einstein-Hilbert action is an interesting (2+1) dimensional modified gravity theory \cite{Berg1,Berg2}. A model which contains
particular higher-derivative terms is known to be renormalizable in
four dimensions \cite{Stelle}. And this implies that the NMG is
super-renormalizable as well as unitary in three dimensions
\cite{Oda1,Naka}. Hence, the NMG can be considered as a physically
interesting toy model of consistent quantum gravity in three
dimensions. Furthermore, in this theory, the graviton gains a mass \cite{Berg1,Naka}. Also, the NMG has non-trivial solutions, such as the
Banados-Teitelboim-Zanelli (BTZ) black hole \cite{Oliva}, Lifshitz black hole
\cite{Beato1}, the Warped-AdS$_{3}$ black hole \cite{Clement}, the New type black hole \cite{kwon}, de
Sitter spacetime \cite{Gab}, AdS3 pp-wave \cite{Beato2,Aliev2}, homogeneous anisotropic of Bianchi type-VI0,
and type-VII0 \cite{Aliev1}. On the other hand, the Hawking radiation of these black holes were investigated by the quantum mechanical tunnelling of the scalar, Dirac particles \cite{gy1,gy2,qi} and vector boson particles\cite{gy3}.

Giving the observational data \cite{ade1,ade2,sper1,sper2}, the
universe had been inflated in early time and since then it keeps
expanding. To investigate the early time inflation of the universe, at first, the standard cosmological models were
used \cite{alan,linde}. However, recently, to understand the problem in both 2+1 and 3+1
dimensional spacetimes, it has been realized that the Dirac field as a gravitational source which cause early time inflationary and late time acceleration is being considered \cite{Saha1,Rib1,Rib2,vakili,Kremer1,GYY,GY,Y}. As the Dirac theory has a vacuum including Zitterbewegung
oscillations between positive and negative energy states and
perfectly explains an interaction of a Dirac particle with an atomic
structure of the materials, it gives a reasonable cosmological
solutions for the early-time inflation and late-time acceleration of
the universe \cite{GYY,GY}. In this motivation, we will couple the Dirac field with the curvature scalar and consider the
Dirac field as a source of the inflation of the universe in the context of the 2+1 dimensional New Massive Gravity by using
the Noether symmetry approach.

This paper is organized as follows. In the following section, we
give the field equations of a theory in which the Dirac field is
minimal coupled to the NMG. In this section, at first, the lagrangian of the system is derived, and, then, using the Noether
symmetry approach the cosmological solutions for this model are found. Finally, we conclude with a summary of the obtained
results. Throughout the paper, we use $c=G=\hbar =1$.

\section{Minimal Coupling Dirac Field to New Massive Gravity} \label{b1}

The action for the NMG is written in the following form \cite{Berg1},
\begin{equation}
S=\int d^{3}x\sqrt{\left\vert g\right\vert }\left[ R-2\Lambda -\frac{1}{m^{2}%
}K\right]  \label{Eqn1}
\end{equation}
where $\Lambda $ is the cosmological parameter, $m$ is the mass of
the graviton and the scalar $K$ is given in terms of Ricci scalar,$R$, and
Ricci tensor, $R_{\mu\nu}$, as follows.
\begin{equation*}
K=R_{\mu \nu }R^{\mu \nu }-\frac{3}{8}R^{2}.
\end{equation*}
Hence, the minimal coupling the NMG with the Dirac field is
\begin{equation}
S=\int d^{3}x\sqrt{\left\vert g\right\vert }\left[ R-2\Lambda -\frac{1}{m^{2}
}K+\frac{i}{2}\left[ \overline{\Psi}\sigma^{\mu }D_{\mu }\Psi
-(\overline{D}_{\mu}\overline{\Psi}) \sigma^{\mu}\Psi
\right]-V(\Psi)\right] \label{Eq2}
\end{equation}
where $V(\Psi )$ represents the self-interaction potential of the
Dirac field, and it depends on only functions of the bilinear $\Psi =\bar{%
\psi}\psi $, $g$ is the determinant of the $g_{\mu \nu
}$ metric tensor, $\psi $ is Dirac spinor with two components, particle and
anti-particle, $\bar{\psi}$ is adjoint of the $\psi $
and $\bar{\psi}=\psi^{\dag }\sigma ^{3}$. Also, $D_{\mu }$=$\partial_{\mu}-\Gamma_{\mu
}(x)$ is the covariant derivative of the Dirac spinor in terms of the spin connection, $\Gamma
_{\mu }(x)$. The spin connections are given as
\begin{equation}
\Gamma_{\mu}(x)=\frac{1}{4}g_{\lambda \alpha }(e_{\nu ,\mu
}^{i}e_{i}^{\alpha }-\Gamma _{\nu \mu }^{\alpha })s^{\lambda \nu
}(x), \label{spin}
\end{equation}%
where $\Gamma _{\nu \mu }^{\alpha }$ is Christoffell symbol, and
$g_{\mu \nu }$ is given in term of triads, $e_{\mu }^{(i)}(x),$ as
follows,
\begin{equation}
g_{\mu \nu }(x)=e_{\mu }^{i}(x)e_{\nu }^{j}(x)\eta _{ij},
\label{met1}
\end{equation}%
where $\mu $ and $\nu $ are curved spacetime indices running from $0$ to $2$. $i$ and $j$ are flat spacetime indices running from $0$ to $2$ and
$\eta _{ij}$ is the $2+1$ dimensional Minkowskian metric with a
signature (1,-1,-1). The spin operators, $s^{\lambda \nu }(x)$, are
given by
\begin{equation}
s^{\lambda \nu }(x)=\frac{1}{2}[\overline{\sigma }^{\lambda }(x),\overline{%
\sigma }^{\nu }(x)],  \label{met2}
\end{equation}%
where $\bar{\sigma}^{\mu}(x)$ are the spacetime dependent Dirac
matrices in the $2+1$ dimensional \cite{sucu1}. Then, using triads, $e_{(i)}^{\mu}(x)$, $\bar{\sigma}^{\mu}(x)$ can be related to the flat spacetime
Dirac matrices, $\overline{\sigma}^{i}$, in the following form
\begin{equation}
\bar{\sigma}^{\mu}(x)=e_{(i)}^{\mu}(x)\bar{\sigma}^{i},
\label{met3}
\end{equation}%
where $\bar{\sigma}^{i}$ are
\begin{equation}
\bar{\sigma}^{0}=\sigma ^{3}\ \ ,\bar{\sigma}^{1}=i\sigma ^{1},\ \bar{\sigma}%
^{2}=i\sigma^{2}.  \label{met4}
\end{equation}
$\sigma^{1}$, $\sigma^{2}$ \ and $\sigma^{3}$ are Pauli matrices
\cite {sucu1}.

To analyse the expansion of the universe in the NMG context, we will consider the spatially flat spacetime background
in the $2+1$ dimensional Friedmann-Robertson-Walker metric as follows,
\begin{equation}
ds^2 = dt^2 - a^2(t)[dx^2 + dy^2],\label{FRW}
\end{equation}
where $a(t)$ is the scale factor of the universe. Hence, in this spacetime background, using the Eq.(\ref{met1}), Eq.(\ref{met2}), Eq.(\ref{met3}),
Eq.(\ref{met4}) and Eq.(\ref{Eq2}), the Lagrangian of the system is written as follows;
\begin{eqnarray}
L=-2 (\dot{a}^{2}+a^{2}\Lambda)
+\frac{\dot{a}^{4}}{6m^{2}a^{2}}+ \frac{i a^2}{2}\left[\overline{\psi}\sigma^{3}\dot{\psi}-\dot{\overline{\psi}}\sigma^{3}\psi\right] -a^{2}V(\Psi). \label{Eqn3}
\end{eqnarray}
Accordingly, from this Lagrangian, the equations of motion for $\psi$, $\overline{\psi}$ and $a$ are
obtained as follow, respectively,
\begin{equation}
\dot{\psi }+H\psi +i\sigma^{3}\psi V^{^{\prime }}=0 \label{Eqn4}
\end{equation}
\begin{equation}
\dot{\overline{\psi }}+H\overline{\psi }-i\overline{\psi}\sigma^{3}V^{^{\prime}}=0 \label{Eqn5}
\end{equation}
\begin{equation}
\frac{\ddot{a}}{a}=-\frac{m^{2}\left(4\rho_{\Lambda}-4\rho_{D}-p_{D}\right)}{2H^{2}-4m^{2}} \label{Eqn6}
\end{equation}
where
\begin{eqnarray}
p_{D}&=&2V^{^{\prime}}\Psi-4V , \quad \rho _{\Lambda }=\Lambda \nonumber \\
\rho_{D}&=&\frac{H^{4}}{4m^{2}}+\frac{V}{2} \label{pres}
\end{eqnarray}
$p_{D}$ and $\rho_{D}$ are pressure and energy density of the Dirac
field, respectively, and $\rho_{\Lambda}$ is vacuum (or dark energy) energy density, and  $H=\dot{a}/a$ is the Hubble parameter. Furthermore, as the dot represents
differentiation with respect to cosmic time $t$, the prime denotes the derivative with respect to $\Psi$.
Moreover, with these equations, using the Hamiltonian constraint equation, $E_L=0$, Friedman equation is obtained as follows;
\begin{eqnarray}
H^2 = \rho_{\Lambda }+\rho_{D}. \label{Eqn7}
\end{eqnarray}
Using the Noether symmetry approach for the Lagrangian Eq.(\ref{Eqn3}), we obtain the
following differential equations by imposing the fact that the
coefficients of $\dot{a}^{4}$, $\dot{a}^{3}\dot{\psi}$, $\dot{a}^{3}\dot{\psi}$ and so on
vanish. The configuration space of the Lagrangian is $Q=(a,\psi_{j},\psi_{j}^\dagger)$, whose tangent
space is $TQ =(a,\psi_{j}, \psi_{j}^\dagger, \dot{a}, \dot{\psi_{j}}, \dot{\psi_{j}^\dagger})$. The existence of Noether symmetry given by $\pounds _{\mathbf{X}} L = 0$
implies the existence of a vector field $\mathbf{X}$ \cite{noo,Demi,Cap1} such that
\begin{eqnarray}
\mathbf{X} &=&\alpha \frac{\partial}{\partial a} + \dot{\alpha} \frac{%
\partial}{\partial \dot{a}}+ \sum_{j=1}^{2}\left(\beta_{j} \frac{\partial}{\partial \psi_{j}} + \dot{%
\beta_{j}} \frac{\partial}{\partial \dot{\psi_{j}}} + \gamma_{j} \frac{%
\partial}{\partial \psi_{j}^\dagger} + \dot{\gamma_{j}} \frac{\partial}{%
\partial \dot{\psi_{j}^\dagger}}\right) ,  \label{vecf}
\end{eqnarray}
where $\alpha, \beta_{j}$ and $\gamma_{j}$ are functions of $a, \psi_{j}$ and $%
\psi_{j}^\dagger$. Hence, the Noether symmetry condition for
Eq.(\ref{Eqn3}) under the $\mathbf{X}$ vector field leads to the following differential equations.
\begin{eqnarray}
2a\frac{\partial \alpha }{\partial a}-\alpha &=&0 \label{m1}
\end{eqnarray}
\begin{eqnarray}
\sum_{j=1}^2\frac{\partial \alpha }{\partial \psi _{i}^{\dagger
}}&=&0 \label{m2}
\end{eqnarray}
\begin{eqnarray}
\sum_{j=1}^2\frac{\partial \alpha }{\partial \psi _{i}}&=&0
\label{m3}
\end{eqnarray}
\begin{eqnarray}
\frac{\partial \alpha }{\partial a} &=&0 \label{m4}
\end{eqnarray}
\begin{eqnarray}
\sum_{j=1}^2\left( \psi _{i}^{\dagger }\frac{\partial \gamma _{i}}{%
\partial a}-\psi _{i}\frac{\partial \beta _{i}}{\partial a}\right)
&=&0 \label{m5}
\end{eqnarray}
\begin{eqnarray}
2\alpha \psi _{i}^{\dagger }+a\beta _{i}-a\sum_{j=1}^2\left( \psi
_{i}\frac{\partial \beta _{i}}{\partial \psi _{j}}-\psi _{i}^{\dagger }\frac{%
\partial \gamma _{i}}{\partial \psi _{j}}\right)  &=&0 \label{m6}
\end{eqnarray}
\begin{eqnarray}
2\alpha \psi _{i}^{\dagger }+a\gamma _{i}+a\sum_{j=1}^2\left( \psi
_{i}\frac{\partial \beta _{i}}{\partial \psi _{j}^{\dagger }}-\psi
_{i}^{\dagger }\frac{\partial \gamma _{i}}{\partial \psi _{j}^{\dagger }}\right)  &=&0 \label{m7}
\end{eqnarray}
\begin{equation}
\alpha \left[ 4\Lambda -2V\right] -aV^{^{\prime }}
\sum_{i=1}^2\epsilon_{i}\left(\gamma_{i}\psi_{i}^{\dagger}+\beta_{i}\psi \right) =0 \label{m8}
\end{equation}

From the Eq.(\ref{m1}), Eq.(\ref{m2}), Eq.(\ref{m3}) and
Eq.(\ref{m4}), we see that $\alpha$=$0$. Furthermore, from the solutions of
Eq.(\ref{m5}),Eq.(\ref{m6}) and Eq.(\ref{m7}), we get the following
results.
\begin{equation*}
\beta _{i}=\kappa \epsilon _{i}\left(\psi_{i}^{\dagger }+\psi
_{i}\right),
\end{equation*}
\begin{equation*}
\gamma _{i}=-\kappa \epsilon _{i}\left(\psi _{i}-\psi_{i}^{\dagger
}\right),
\end{equation*}
where $\kappa$ is a constant and $\epsilon_{i}$ is
\begin{equation*}
\epsilon_{i}=\left\{
\begin{array}{c}
1~~ ~~\textstyle{for}~~ i=1\;\cr -1~~ \textstyle{for}~~ i=2.\;
\end{array}
\right.
\end{equation*}
Finally, the expression of the potential is obtained from the equation Eq.(\ref{m8});
\begin{equation}
V=V_{0}, \label{pot2}
\end{equation}
and substituting this potential expression to Eq.(\ref{Eqn4}) and Eq.(\ref{Eqn5}), we get the following equation,
\begin{eqnarray}
\dot{\Psi} + 2 \frac{\dot{a}}{a}\Psi=0.  \label{dm1}
\end{eqnarray}
And this equation gives us the following relation,
\begin{eqnarray}
\Psi = \frac{\Psi_0}{a^2},  \label{dm2}
\end{eqnarray}
where $\Psi_0$ is a integration constant. Then the field equations Eq.(\ref{Eqn6}) and Eq.(\ref{Eqn7}) are reduced to the following expressions, respectively;
\begin{equation}
\frac{\ddot{a}}{a}\left[\frac{2}{m^{2}}\left(\frac{\dot{a}}{a}\right)^{2}-4\right]=-4\Lambda+\frac{1}{m^{2}}\left(\frac{\dot{a}}{a}\right)^{4}-2V_{0} \label{Eqn6s}
\end{equation}
and
\begin{eqnarray}
\left(\frac{\dot{a}}{a}\right)^{2} = \Lambda+\frac{1}{4m^{2}}\left(\frac{\dot{a}}{a}\right)^{4}+\frac{V_{0}}{2}. \label{Eqn7s}
\end{eqnarray}
These equations have four physical solutions. The first two solutions are given as follow,
\begin{equation}
a_{1,2}(t)=e^{\mp \left( t-t_{0}\right) K_{1}}  \label{dSit}
\end{equation}
where $K_{1}=\sqrt{2m\sqrt{m^{2}-\lambda}+2m^{2}}$ and $\lambda=\Lambda+V_{0}/2$. Under the $m^{2}\geq|\lambda|$ and $\lambda>0$ conditions, they correspond to the Sitter space-time \cite{Gab}. The rest two solutions are given as;
\begin{equation}
a_{3,4}(t)=e^{\mp i \left( t-t_{0}\right) K_{2}}, \label{osi}
\end{equation}
where $K_{2}=\sqrt{2m\sqrt{m^{2}-\lambda}-2m^{2}}$, and they correspond to an oscillation universe (or cyclic universe) model under the $m^{2}\geq|\lambda|$ and the $\lambda<0$ (i.e. anti-de Sitter space) conditions \cite{Barrow}.

On the other hand, using the Eq.(\ref{pres}) and Eq.(\ref{pot2}), the pressure of the Dirac field can be calculated as follows;
\begin{equation}
p_{D}=-V_{0}. \label{presx}
\end{equation}
This result implies that the Dirac field has negative pressure for $V_{0}>0$. This means that the Dirac field behaves like a dark energy, and
using Eq.(\ref{pres}), the energy density of the Dirac field for Eq.(\ref{dSit}) and for Eq.(\ref{osi}) is obtained as follows;
\begin{equation}
\rho_{D}=\left[\sqrt{m^{2}-\lambda}+m\right]^{2}+\frac{V_{0}}{2}, \label{pres11}
\end{equation}
\begin{equation}
\rho_{D}=\left[\sqrt{m^{2}-\lambda}-m\right]^{2}+\frac{V_{0}}{2}. \label{pres22}
\end{equation}
Using $\Lambda=\rho_{\Lambda}$ in Eqs.(\ref{pres11}) and (\ref{pres22}), we see
that the energy density of the Dirac field depends on the graviton mass and the vacuum energy density. All these results indicate a Dirac field of the characteristic features of the dark energy in the context of the NMG.

\section{Concluding remarks}\label{conc}

The study of cosmological models in the gravitation theories with
the $2+1$ dimensional provide for mathematical simplicity than that of the $3+1$ dimensional theories. In the present study, we couple the Dirac field with the NMG. Also, the effects of the
Dirac field on the evolution of the universe are probed. In contrast to standard three-dimensional Einstein gravity, we see that the Dirac field has a negative pressure. This result indicates that the Dirac field plays role of the dark energy in the evolution of the universe. Hence, in the context of the NMG, the Dirac field provides explanation for the early-time inflation period of the universe. Furthermore, we have also seen that the Dirac field can lead to an oscillating universe in time.

Using these four solutions with Hubble parameter, $H$=$\dot{a}/a$, and deceleration parameter, $q$=$-a\ddot{a}/\dot{a}^{2}$, we can analyze the evolution of the universe in the context of the NMG as follows:
\begin{itemize}
\item The Hubble and the deceleration parameters become as $H$=$K_{1}$ and $q$=$-1$, respectively, for $a_{1}(t)$=$exp[K_{1}(t-t_{0})]$. This case corresponds to an accelerating expansion phase, i.e. the early-time inflation epoch (i.e. $H>0$ and $q<0$).
\item When $a_{1}(t)$=$exp[-K_{1}(t-t_{0})]$, the Hubble parameter and the deceleration parameter become as $H$=$-K_{1}$ and $q$=$-1$, respectively, and this case represents a contracting expansion phase (i.e. $H<0$ and $q<0$).
\item Using the Eqs.(\ref{osi}), Hubble and deceleration parameters obtained as $H$=$\mp i K_{2}$ and $q$=$-1$. The signature (+) corresponds to an accelerating expansion phase (i.e. $H>0$ and $q<0$) and the signature (-) corresponds to a contracting expansion phase (i.e. $H<0$ and $q<0$). Furthermore, the imaginary Hubble parameter point out that the universe is oscillating in time.
\end{itemize}

\section*{Conflict of Interests}
The authors declare that there is no conflict of interests regarding the publication of this paper.

\section*{Acknowledgments}

This work was supported by the Scientific Research Projects Unit of Akdeniz University.



\section*{References}

\begin{thebibliography}{99}

\bibitem{Berg1} E.A. Bergshoeff, O. Hohm, P.K. Townsend, "Massive Gravity in Three Dimensions", \textit{Phys. Rev. Lett.} \textbf{102} 201301 (2009).
\bibitem{Berg2} E.A. Bergshoeff, O. Hohm, P.K. Townsend, "More on massive 3D gravity", \textit{Phys. Rev. D} \textbf{79} 124042 (2009).
\bibitem{Stelle} K.S. Stelle, "Renormalization of higher-derivative quantum gravity", \textit{Phys. Rev. D} \textbf{16} 953 (1977).
\bibitem{Oda1} I. Oda, "Renormalizability of Massive Gravity in Three Dimensions", \textit{JHEP} \textbf{05} 064 (2009).
\bibitem{Naka} M. Nakasone and I. Oda, "On Unitarity of Massive Gravity in Three Dimensions", \textit{Prog.Theor.Phys} \textbf{121} 1389 (2009).
\bibitem{Oliva} J. Oliva, D. Tempo and R. Troncoso, "Three-dimensional black holes, gravitational solitons, kinks and wormholes for BHT massive gravity", \textit{JHEP} \textbf{07} 011 (2009).
\bibitem{Beato1} E. Ayon-Beato, A. Garbarz and M. Hassaine, "Lifshitz black hole in three dimensions", \textit{Phys. Rev. D} \textbf{80} 104029 (2009)
\bibitem{Clement} Clement, G. "Warped AdS3 black holes in new massive gravity", \textit{Class. Quantum Grav.}\textbf{26}, 105015 (2009).
\bibitem{kwon} Y. Kwon, S. Nam, J. Park and S. Yi, "Quasinormal modes for new type black holes in new massive gravity", \textit{Class. Quantum Grav.}, 28, 145006, (2011).
\bibitem{Gab} G. Gabadadze, G. Giribet and A. Iglesias, "New Massive Gravity on de Sitter Space and Black Holes at the Special Point", arXiv:1212.6279 [gr-qc] (2012).
\bibitem{Beato2} E. Ayon-Beato, G. Giribet and M. Hassaine, "Bending AdS waves with new massive gravity", \textit{JHEP} \textbf{05} 029 (2009)
\bibitem{Aliev2} H. Ahmedov and A.N. Aliev, "The general type N solution of new massive gravity", \textit{Phys. Lett. B} \textbf{694} 143 (2010).
\bibitem{Aliev1} H. Ahmedov and A.N. Aliev, "Exact Solutions in 3D New Massive Gravity" \textit{Phys. Rev. Lett.} \textbf{106} 021301 (2011)
\bibitem{gy1} G. Gecim and Y. Sucu, "Tunnelling of relativistic particles from new type black hole in new massive gravity", \textit{JCAP}, \textbf{02}, 023 (2013)
\bibitem{gy2} G. Gecim and Y. Sucu, "Dirac and scalar particles tunnelling from topological massive warped-AdS3 black hole", Astrophys Space Sci. \textbf{357}, 105 (2015).
\bibitem{gy3} G. Gecim and Y. Sucu, "Massive vector bosons tunnelled from the (2+1)-dimensional black holes", Eur. Phys. J. Plus \textbf{132}, 105 (2017).
\bibitem{qi} D.J. Qi, "Fermions Tunneling Mechanism for a New Class of Black Holes in EGB Gravity and Three-Dimensional Lifshitz Black Hole" : \textit{Int. J. Theor. Phys.}, \textbf{52}, 345, (2013).
\bibitem{ade1} P.A.R. Ade, et al., "Detection of B-Mode Polarization at Degree Angular Scales by BICEP2", \textit{Phys. Rev. Lett.} \textbf{112}  241101 (2014)
\bibitem{ade2} P.A.R. Ade, et al., "Planck 2013 results. XVI. Cosmological parameters", \emph{Astronomy and Astrophysics}, vol. 571, A16, [arXiv:1303.5076], 2014.
\bibitem{sper1} D.N. Spergel, et al., "First-year Wilkinson Microwave Anisotropy Probe (WMAP) observations: Determination of cosmological parameters", \textit{Astrophys. J. Suppl.} \textbf{148}  175 (2003)
\bibitem{sper2} D.N. Spergel, et al., "Three-year Wilkinson Microwave Anisotropy Probe (WMAP) observations: Implications for cosmology", \textit{Astrophys. J. Suppl.} \textbf{177}  377 (2007)
\bibitem{alan} A.H. Guth, "Inflationary universe: A possible solution to the horizon and flatness problems", \textit{Phys. Rev. D}\textbf{23} 347 (1981)
\bibitem{linde} A.D. Linde, "Chaotic Inflation", \textit{Phys. Lett. B}\textbf{129} 177 (1983)
\bibitem{Saha1} B. Saha, "Spinor field in a Bianchi type-1 universe: Regular solutions", \textit{Phys. Rev. D} \textbf{64} 123501 (2001)
\bibitem{Rib1} M.O. Ribas, M.O. Devecchi, G.M. Kremer, "Fermions as sources of accelerated regimes in cosmology", \textit{Phys. Rev. D}\textbf{72}, 123502 (2005)
\bibitem{Rib2} M.O. Ribas, G.M. Kremer, "Cosmological model with non-minimally coupled fermionic field", \textit{Europhys Lett.}\textbf{81}, 19001 (2008).
\bibitem{vakili} B. Vakili, S. Jalalzadeh, H.G. Sepangi, "Classical and quantum spinor cosmology with signature change", \textit{J. Cosmol. Astropart. Phys.} \textbf{05}, 006 (2008)
\bibitem{Kremer1} R.C. De Souza, G.M. Kremer, "Noether symmetry for non-minimally coupled fermion fields", \textit{Class. Quantum Grav.}\textbf{25}, 225006 (2008)
\bibitem{GYY} G. Gecim, Y. Kucukakca and Y. Sucu, "Noether Gauge Symmetry of Dirac Field in (2+1)-Dimensional Gravity", \emph{Advances in High Energy Physics}, vol. 2015, Article ID. 567395, 2015.
\bibitem{GY} G. Gecim and Y. Sucu, "Dirac Field as a Source of the Inflation in $2+1$ Dimensional Teleparallel Gravity", \emph{Advances in High Energy Physics}, vol. 2017, Article ID. 2056131, 2017.
\bibitem{Y} Y. Kucukakca, "Teleparallel dark energy model with a fermionic field via Noether symmetry", \emph{The European Physical Journal C}, 74, 3086, 2014.
\bibitem{sucu1} Y. Sucu and N. Unal, "Exact solution of Dirac equation in 2+1 dimensional gravity", \emph{Journal of Mathematical Physics}, vol. 48, no. 5, Article ID 052503, 2007.
\bibitem{noo} E. Noether, \textit{Nachr. d. K\"{o}nig. Gesellsch.d. Wiss. zu G\"{o}ttingen, Math-phys. Klasse}, 235-257, 1918.
\bibitem{Demi} M. Demianski, et al., "Scalar field, nonminimal coupling, and cosmology", \emph{Physical Review D}, vol. 44, no. 10, pp. 3136-3146, 1991.
\bibitem{Cap1} S. Capozziello and R. de Ritis, "Relation between the potential and nonminimal coupling in inflationary cosmology", \emph{Physics LettersA}, 177, 1-7, 1993.
\bibitem{Barrow} J.D. Barrow and M.P. Dabrowski, "Oscillating Universes", \emph{Mon. Not. R. Astron. Soc.}, 275, 850, 1995.


\end{thebibliography}
\end{document}